\documentclass[a4paper,10pt,twoside]{cpc-hepnp}
\usepackage{CJK,upgreek,fancyhdr}
\usepackage[utf8]{inputenc}
\usepackage{multicol}
\usepackage{booktabs}
\usepackage{amssymb}
\usepackage{mathrsfs}
\usepackage{bbm}
\usepackage{amscd}
\usepackage{lastpage}
\usepackage[tbtags]{amsmath}
\usepackage{graphicx}
\usepackage{dcolumn}
\usepackage{bm}
\usepackage{hyperref}
\usepackage{siunitx}
\usepackage{slashed}
\usepackage{subfigure}
\usepackage{natbib}

\begin{document}


\fancyhead[c]{\small Chinese Physics C~~~Vol. 40, No. 3 (2016) 033103}
\fancyfoot[C]{\small 033103-\thepage}

\footnotetext[0]{}

\title{Efficient Numerical Evaluation of Feynman Integrals
\thanks{
This work was supported by the Natural Science Foundation of China under Grant No. 11305179 and No. 11475180; 
and by Youth Innovation Promotion Association, CAS; 
and by the IHEP Innovation Grant under contract number Y4545170Y2,
and is partially supported by the State Key Lab for Electronics and Particle Detectors;
and by the Open Project Program of State Key Laboratory of Theoretical Physics, 
Institute of Theoretical Physics, Chinese Academy of Sciences, China (No.Y4KF061CJ1);
J.W. was supported by the Cluster of Excellence 
{\it Precision Physics, Fundamental Interactions and Structure of Matter} (PRISMA-EXC 1098). 
}
}
\author{%
  Zhao Li$^{1,2:1)}$\email{zhaoli@ihep.ac.cn}%
\quad Jian Wang$^{3;2)}$\email{jian.wang@uni-mainz.de}%
\quad Qi-Shu Yan$^{4,5;3)}$\email{yanqishu@ucas.ac.cn}%
\quad Xiaoran Zhao$^{1,4;4)}$\email{zhaoxiaoran13@mails.ucas.ac.cn}
}

\maketitle

\address{
$^1$ Institute of High Energy Physics, Chinese Academy of Sciences, Beijing 100049, P.R. China\\
$^2$ State Key Laboratory of Theoretical Physics, Institute of Theoretical Physics, Chinese Academy of
Sciences, Beijing 100190, P.R. China\\
$^3$ PRISMA Cluster of Excellence \& Mainz Institute for Theoretical Physics, Johannes Gutenberg
University, D-55099 Mainz, Germany\\
$^4$ School of Physics Sciences, University of Chinese Academy of Sciences, Beijing 100049, P.R. China \\
$^5$ Center for High-Energy Physics, Peking University, Beijing, 100871, P.R. China \\
}

\begin{abstract}
Feynman loop integrals are a key ingredient for the calculation of higher order radiation effects, 
and are responsible for reliable and accurate theoretical prediction.
We improve the efficiency of numerical integration in sector decomposition 
by implementing a quasi-Monte Carlo method associated with the CUDA/GPU technique.
For demonstration we present the results of several Feynman integrals 
up to two loops in both Euclidean and physical kinematic regions 
in comparison with those obtained from FIESTA3.
It is shown that both planar and non-planar two-loop master integrals in the physical kinematic region can be evaluated
in less than half a minute with $\mathcal{O}(10^{-3})$ accuracy,
which makes the direct numerical approach viable for precise investigation of 
higher order effects in multi-loop processes, e.g. the next-to-leading order QCD effect 
in Higgs pair production via gluon fusion with a finite top quark mass.

\end{abstract}



\footnotetext[0]{\hspace*{-3mm}\raisebox{0.3ex}{$\scriptstyle\copyright$}2013
Chinese Physical Society and the Institute of High Energy Physics
of the Chinese Academy of Sciences and the Institute
of Modern Physics of the Chinese Academy of Sciences and IOP Publishing Ltd}%

\begin{multicols}{2}
\section{Introduction}
The scalar particle predicted by the Standard Model, the Higgs boson, 
has been finally discovered at the CERN Large Hadron Collider (LHC)\cite{Aad:2012tfa,Chatrchyan:2012ufa}.
This milestone has inspired various exciting investigations on the further details of the Higgs boson and related research.
Recently the launch of LHC RunII at 13~TeV collision energy brings physics exploration to a new era.
With the highest ever center-of-mass energy and luminosity, 
many scattering processes that potentially answer some fundamental physics questions will be able to reach  
an accuracy in the percent range or better,
while the appropriate analysis on high precision data demands that
the uncertainties of theoretical prediction reach the same accuracy.
In order to achieve this, higher order QCD effects must be included in theoretical predictions.
For instance, state-of-the-art investigations on Higgs inclusive production have explored 
next-to-next-to-next-to-leading order (NNNLO) effects for the quark pair annihilation initial state \cite{Anzai:2015wma} 
and for the gluon fusion initial state \cite{Anastasiou:2015ema}.
Meanwhile, next-to-next-to-leading order (NNLO) theoretical predictions have been provided for 
Higgs pair production \cite{deFlorian:2013jea} 
and the associated production of Higgs with jet\cite{Chen:2014gva,Boughezal:2015dra,Boughezal:2015aha} 
or vector boson \cite{Brein:2003wg,Brein:2011vx,Ferrera:2011bk,Ferrera:2014lca}.
As the accumulated luminosity at the LHC increases, the investigation of higher order QCD effects  
will be wanted for more processes, e.g. top quark production and jet cross sections.
In the higher order effects, one of the most important ingredients is virtual correction,
which always relies on evaluation of Feynman loop integrals.

After decades of effort, various algorithms have been proposed for evaluating Feynman loop integrals,  
including both analytical and numerical approaches.
The analytical approaches can provide explicit expressions for Feynman integrals, 
and can further reveal significant physical characteristics. 
However, when complicated processes are encountered, 
it becomes difficult to obtain analytical solutions, while the numerical approaches 
can solve more challenging problems in spite of a heavy burden of evaluation time.
Sector decomposition, one of the numerical algorithms, was introduced as a systematic approach 
by Binoth and Heinrich\cite{Binoth:2000ps,Binoth:2003ak}.
Following a certain choice of decomposition strategies, 
this algorithm divides the domain of loop integration into sectors.
In each individual sector, proper transformation of integration variables is
performed to explicitly reveal the ultraviolet (UV) and infrared (IR) singularities.
Ultimately the coefficients of a Laurent series in $\epsilon$ of the Feynman integral can be evaluated numerically.
Initially, sector decomposition was implemented for the Feynman integral 
in the Euclidean kinematic region\cite{Binoth:2000ps,Binoth:2003ak,Heinrich:2004iq}, 
where the Cauchy singular integral can be avoided.
Later, inspired by Nagy and Soper\cite{Nagy:2003qn,Nagy:2006xy},
integration contour deformation was proposed \cite{Binoth:2005ff} 
as a systematic scheme to extend sector decomposition to the physical kinematic region.

Several programs \cite{Bogner:2007cr,Smirnov:2013eza,Borowka:2015mxa} have implemented 
the sector decomposition algorithm for the numerical evaluation of Feynman loop integrals.
Normally they use Monte Carlo (MC) integration methods, which have been widely used in high energy physics research. 
For instance, Vegas\cite{Lepage:1977sw}, an adapative Monte Carlo method, 
can achieve an integration convergence rate of ${\mathcal O}(1/\sqrt{n})$.
In this paper, we implement the quasi-Monte Carlo (QMC) method 
for the numerical evaluation of the integrals in sector decomposition.
As a better choice, QMC can have a convergence rate close to ${\mathcal O}(n^{-1})$ for differentiable integrands.
Furthermore, we adopt the technique of CUDA/GPU to improve the performance of numerical evaluation.

This paper is organised as follows. In Section II we review the sector decomposition algorithm, then
Section III gives a brief description of the QMC integration method.
In Section IV we compare the performance of our program with FIESTA\cite{Smirnov:2013eza}.
Our conclusions are then presented in the final section.

\section{sector decomposition}

Generically an $L$-loop Feynman integral has the following representation:
\begin{equation}
I=\int\left(\prod_{l=1}^{L}\dfrac{\mathrm{d}^D k_l}{i\pi^{\dfrac{D}{2}}}\right)
\prod_{j=1}^{N}\dfrac{1}{(q_j^2-m_j^2+i\varepsilon)^{\nu_j}},
\end{equation}
where $D=4-2\epsilon$; $q_j$ is the momentum of relevant internal propagator, 
and is a linear combination of the loop momenta $\{k_l\}$ and
external momenta; $m_j$ is the mass of the relevant internal propagator; and
$\nu_j$ is the power of the corresponding propagator. 

After Feynman parameterisation and integration over the loop momenta, one can obtain
\begin{align}
  \begin{split}
I=&(-1)^{N_{\nu}}\dfrac{\Gamma(N_{\nu}-LD/2)}{\prod^{N}_{j=1}\Gamma(\nu_j)}
\int^{\infty}_0\cdots\int^{\infty}_0\left(\prod_{l=1}^{N}\mathrm{d}x_l~ x_l^{\nu_l-1}\right)\\
&\delta\left(1-\sum^{N}_{l=1}x_l\right)
\dfrac{U^{N_{\nu}-(L+1)D/2}(x_1,\cdots,x_N)}{F^{N_\nu-LD/2}(x_1,\cdots,x_N)},
\end{split}
\end{align}
where $N_\nu=\sum_{j=1}^{N}\nu_j$, and $U$ and $F$ are polynomials of $\{x_l\}$ 
and can be straightforwardly derived from the momentum representation, 
or constructed from the topology of the corresponding Feynman graph \cite{Bogner:2010kv}.

Further treatment of the Feynman integral requires careful consideration 
since $U$ and $F$ can vanish when some of $\{x_l\}$ approach zero,
which may be related to UV or IR divergence.
Direct numerical integration is impossible for divergent integrals.
A sector decomposition algorithm is designed to systematically extract the divergence, 
and is briefly described as follows\cite{Heinrich:2008si}.

Firstly, the integration domain is equally split into $N$ sub-domains, which are called primary sectors:
\begin{equation}
  \int^{\infty}_0\mathrm{d}^{N}x=\sum_i^{N}\int^{\infty}_0\mathrm{d}^{N}x\ \prod^{N}_{\substack{j=1\\j\neq i}}\theta(x_i-x_j).
\end{equation}
Then in the $i$-th sector we implement the variable transformation, 
\begin{align}
  x_j=\begin{cases}
        x_i t_j &j < i,\\
        x_i &j = i,\\
        x_i t_{j-1} &j > i.\\
  \end{cases}
\end{align}
Thereafter $x_i$ integration is performed associated with the step function, 
and now the Feynman integral can be expressed as
\begin{equation}
I=(-1)^{N_{\nu}}\dfrac{\Gamma(N_{\nu}-LD/2)}{\prod^{N}_{j=1}\Gamma(\nu_j)}\sum_l^{N}I_l,
\end{equation}
where
\begin{equation}
I_l=\int^{1}_{0}\cdots\int^{1}_{0}\left(\prod^{N-1}_{j=1}\mathrm{d}t_j~ t_j^{\nu_j-1}\right)
\dfrac{U_l^{N_{\nu}-(L+1)D/2}(t_1,\cdots,t_{N-1})}{F_l^{N_{\nu}-LD/2}(t_1,\cdots,t_{N-1})}.
\end{equation}
Obviously for any given primary sector $I_l$, 
the domain of integration is an $(N-1)$-dimensional unit cube. 

Next, following an iterative decomposition strategy\cite{Bogner:2007cr} 
or geometric strategy \cite{Kaneko:2009qx,Kaneko:2010kj},
each primary sector is finally divided into some subsectors $\{I_{la}\}$ 
so that in any subsector polynomials $U_l$ and $F_l$ can be factorised into the form
\begin{align}
  U_l &= C_{la} (1+H_{la}(t_1,\cdots,t_{N-1}))\prod_{j=0}^{N-1}t_j^{b_{laj}},\\
  F_l &= C^{\prime}_{la}(1+H^{\prime}_{la}(t_1,\cdots,t_{N-1})) \prod_{j=0}^{N-1}t_j^{b^{\prime}_{laj}},
\end{align}
after proper variable transformation.
In each subsector, the new variables and Jacobian generated 
by the transformation are required to be monomials of original variables,
and meanwhile the transformation projects 
the domain of integration to the $(N-1)$-dimensional unit cube.
In above expressions, $b_{laj}$ and $b^{\prime}_{laj}$ are non-negative integers. 
$H_{la}$ and $H^{\prime}_{la}$ are polynomials of $\{t_j\}$ 
such that $H_{la}(0,\cdots,0)=0$ and $H^{\prime}_{la}(0,\cdots,0)=0$. 

Now the primary sector becomes the combination of subsectors,
\begin{align}
  \begin{split}
  I_l =&\sum_{a=1}^{m} D_{la} \int_0^1\cdots\int_0^1\left(\prod_{j=1}^{N-1}\mathrm{d} t_j~
  t_j^{\alpha_{laj}+\beta_{laj}\epsilon}\right)\\
&\dfrac{(1+H_{la}(t_1,\cdots,t_{N-1}))^{N_\nu-(L+1)D/2}}{(1+H^{\prime}_{la}(t_1,\cdots,t_{N-1}))^{N_\nu-LD/2}},
\end{split}
\end{align}
where the powers of $t_j$ are collected into $\alpha_{laj}+\beta_{laj}\epsilon$, 
and $D_{la}$ contains the coefficients from the Jacobian and $C_{la}$ and $C^{\prime}_{la}$.

In the practical evaluation of Feynman integrals, we will adopt the geometric strategy 
since it can be guaranteed to succeed and results in the smallest number of subsectors.

After sector decomposition, the singularities in the Feynman integral have been collected 
into the regulators in the form of $t^{\alpha+\beta\epsilon}$, which can explicitly present
the pole of the integral by using a Laurent series or integration by parts (IBP).
Without loss of generality, we rewrite the integral with a certain regulator as 
\begin{equation}
\mathcal{I}=\int_0^1\mathrm{d} t~ t^{\alpha+\beta\epsilon} f(t,\epsilon),
\end{equation}
where $f(0,\epsilon)$ is non-zero and finite.
Then if $\alpha \le -1$, the above integral contains a singularity on the lower bound.
By expanding $f(t,\epsilon)$ into a Laurent series around $t=0$, the singularity can be explicitly extracted as
\begin{equation}
\mathcal{I}=\sum_{p=0}^{|\alpha|-1}\dfrac{1}{\alpha+p+1+\beta\epsilon}\dfrac{f^{(p)}(0,\epsilon)}{p!}
  +\int_0^1\mathrm{d}t~t^{\alpha+\beta\epsilon}r(t,\epsilon),
\end{equation}
where
\begin{equation}
r(t,\epsilon)=f(t,\epsilon)-\sum_{p=0}^{|\alpha|-1}f^{(p)}(0,\epsilon)\dfrac{t^p}{p!}.
\end{equation}
Thereafter, the integrands can be expanded by small $\epsilon$, 
and the coefficients of the Laurent series in $\epsilon$ can be evaluated numerically order by order.
However, numerical evaluation of $r(t)$ may suffer numerical instability from large number cancellation. 
An alternative approach to the pole extraction that can avoid this problem is IBP, 
\begin{align}
  \begin{split}
\int_0^{1}\mathrm{d}t~ t^{\alpha+\beta\epsilon}f(t,\epsilon)
=&\dfrac{1}{\alpha+\beta\epsilon+1}f(1,\epsilon)-\\
&\int_0^{1}\mathrm{d}t
\dfrac{\partial f(t,\epsilon)}{\partial t}\dfrac{t^{\alpha+\beta\epsilon+1}}{\alpha+\beta\epsilon+1}.
\end{split}
\end{align}
It can be seen that the power of $t$ increases by one.
Therefore, by repeating the above IBP formula enough times, 
the power of $t$ will not generate singularity, 
and then the numerical approach can be implemented to evaluate the coefficients of the Laurent series in $\epsilon$.

However, occasionally the integral contains Cauchy singularities in the physical kinematic region.
In this case, the sign of $F$ cannot be guaranteed definite, 
so the Cauchy singular Feynman integral is only valid under a proper contour 
according to the conventional $i\varepsilon$ prescription of the Feynman propagators. 
Practically such an infinitesimal shifted contour will sabotage the stability of numerical integration.
Fortunately an interesting prescription of contour deformation has been proposed 
\cite{Anastasiou:2007qb}
\begin{equation}
  z_i(\vec{t})=t_i-i \lambda_i t_i(1-t_i)\dfrac{\partial F(\vec{t})}{\partial t_i},
\end{equation}
where an appropriate choice of $\lambda_i$ can guarantee the sign of ${\rm Im}(F(\vec{z}))$ is always negative as required conventionally.

\section{Quasi-Monte Carlo\label{QMCalgo}}

After the implementation of the sector decomposition algorithm reviewed in Section II, 
the Feynman integral is expressed as a Laurent series in $\epsilon$.
The coefficients of the series are composed of convergent integrals, which can be numerically evaluated.
A one-dimensional integral can be easily evaluated by a numerical approach such as the trapezoidal rule, 
while numerical evaluation of multi-dimensionals integral is usually much more difficult.

For instance, an $s$-dimensional integral can be written as
\begin{align}
  I_s(f)=\int_{[0,1]^s}\mathrm{d}^s{x}f(\vec{x}).
\end{align}
Given a predefined set of $n$ points 
$\{\vec{x}_i |~ \vec{x}_i \in [0,1]^s;~ i = 0,\dots,n-1 \}$, 
the above integral can be estimated by 
\begin{align}
  Q_{n,s}(f)=\dfrac{1}{n}\sum_{i=0}^{n-1}f(\vec{x}_i) \approx I_s(f).
\label{QMCestimation}
\end{align}
This method is called the quasi-Monte Carlo method, and the point set is called the quasi-Monte Carlo
rule\cite{ANU:8877392}.
Conventionally two families of QMC rules attract most interest.  One consists of digital nets and digital sequences, while the other is the lattice rule. 
In this paper we adopt the rank-1 lattice rule (R1LR) defined by\cite{ANU:8877392}
\begin{align}
  \vec{x}_i=\left \{\dfrac{i\vec{z}}{n}\right\},~ i=0,\dots,n-1,
  \label{rankonerule}
\end{align}
where $\vec{z}$, known as the generating vector, 
is an $s$-dimensional integer vector. The integer components of $\vec{z}$ are all relatively prime to $n$. 
The braces around the vector in Eq. (\ref{rankonerule}) means only the fractional part of each component in the vector is taken.

The previous R1LR will result in a biased estimation, 
since it is fully deterministic. To achieve an unbiased result, we need to introduce appropriate randomisation.
For R1LR, we can use the simplest form of randomisation, called shifting, which will yield so-called shifted lattice rule. 
The QMC algorithm utilising random shifted R1LR is explained as below\cite{ANU:8877392}.
\begin{enumerate}
  \item A set of $m$ independent random vectors called shifts, $\{ \vec{\Delta}_1,\cdots,\vec{\Delta}_m \}$,
        is generated with uniform distribution in $[0,1]^s$.
  \item For each shift, the integral estimation in Eq. (\ref{QMCestimation}) 
        is repeated to obtain a set of $m$ integral estimations,
        $\{ Q_{n,s,1}(f),\cdots,Q_{n,s,m}(f)\}$, where
        \begin{align}
          Q_{n,s,k}(f)=\dfrac{1}{n}\sum_{i=0}^{n-1}f\left(\left\{\dfrac{i\vec{z}}{n}+\vec{\Delta}_k\right\}\right), \quad k = 1,\dots,m.
        \end{align}
  \item Then the average of these $m$ integral estimations
        \begin{align}
          \bar{Q}_{n,s,m}(f)=\dfrac{1}{m}\sum_{k=1}^{m}Q_{n,s,k}(f),
        \end{align}
        is finally taken as an unbiased approximation of the integral $I_s(f)$.
  \item Furthermore, an unbiased estimation
        for the mean-square error of $\bar{Q}_{n,s,m}(f)$ can be obtained by
        \begin{align}
          \dfrac{1}{m(m-1)}\sum_{k=1}^{m}(Q_{n,s,k}(f)-\bar{Q}_{n,s,m}(f))^2.
        \end{align}
  \end{enumerate}

The above algorithm improves the practicability of the R1LR QMC method. 
Moreover, in the case that the integrand $f$ is a 1-periodic function 
and all partial derivatives of $f$ exist, 
the convergence rate can be improved to $O(n^{-1})$ \cite{Kuo2003301,Dick2004493}
if the generating vector is obtained by the component-by-component construction.
However, in practice even when the integrand $f$ is a non-periodic function, 
one can implement the transformation $x_i=3y_i^2-2y_i^3$ to obtain a periodic integrand as below:
\begin{align}
I_s(f) &= \int_{[0,1]^s}\mathrm{d}^s x f(\vec{x}) \nonumber\\
  &=\int_{[0,1]^s}\mathrm{d}^s y f(\vec{x}(\vec{y}))\prod_{i=1}^{s}(6y_i(1-y_i))\\
  &=\int_{[0,1]^s}\mathrm{d}^s y g(\vec{y}),\nonumber
\end{align}
where it can be seen that $g(\vec{y})=0$ once $\vec{y}$ reaches boundary of $[0,1]^s$, as long as $f$ is bounded at the edge.

Beside the $O(n^{-1})$ convergence rate, the shifted R1LR QMC method has some intrinsic advantages.
In the shifted R1LR QMC method the lattice rule is deterministic 
and therefore the complexity of random number generation only depends on the number of shifts.
By contrast, in the MC method, since the evaluation points are independent random vectors,
a large amount of pseudo random number generation consumes much more of the GPU resources. 
Besides, since the QMC method is a non-adaptive method, it can easily deal with  integrals in complex space, 
which is inevitable for Feynman integrals in the physical kinematic region. 
However, for adaptive algorithms, e.g. Vegas, it is difficult to define an appropriate rule to handle such integrals.

\section{Numerical Results}

In this section, we present some numerical results for several Feynman integrals up to two loops
for certain choices of kinematic parameter as a demonstration.
In the Euclidean kinematic region we show the evaluation of massless scalar double box diagrams,
and in the physical kinematic region we take several master integrals from the investigation of Higgs pair production via gluon fusion.
The Higgs and top quark masses are chosen as $M_H=125$ GeV and $M_t=172$ GeV \cite{Agashe:2014kda}.
By comparing\footnote{
Our program was deployed with NVIDIA Tesla K20 GPU, while
FIESTA3 used four cores of Intel Core i7 3770 CPU (3.4GHz).
} with FIESTA3\cite{Smirnov:2013eza} using the Vegas algorithm \cite{Hahn:2004fe,Lepage:1977sw} as a benchmark of CPU efficiency, 
we illustrate the improvement in the efficiency of numerical evaluation of Feynman integrals.
The sample codes for the Feynman integrals in this section 
were generated by MIRACLE, which is a general purpose package in preparation.

\subsection{One-loop Feynman integral in physical kinematic region}
The leading order (LO) contribution to Higgs pair production via gluon fusion contains the one-loop box Feynman diagram
as shown in Fig. \ref{boxhh}.
For the evaluation of this diagram, the most complicated master integral is
\begin{align}
  \begin{split}
I_A=&\int\dfrac{\mathrm{d}^D k}{(2\pi)^D}\dfrac{1}{(-k^2+M_t^2+i\varepsilon)[-(k+p_1)^2+M_t^2+i\varepsilon]}\\
&\times\dfrac{1}{[-(k-p_2)^2+M_t^2+i\varepsilon]
[-(k+p_1-p_4)^2+M_t^2+i\varepsilon]}.
\end{split}
\end{align}

\begin{center}
  \includegraphics[width=250px]{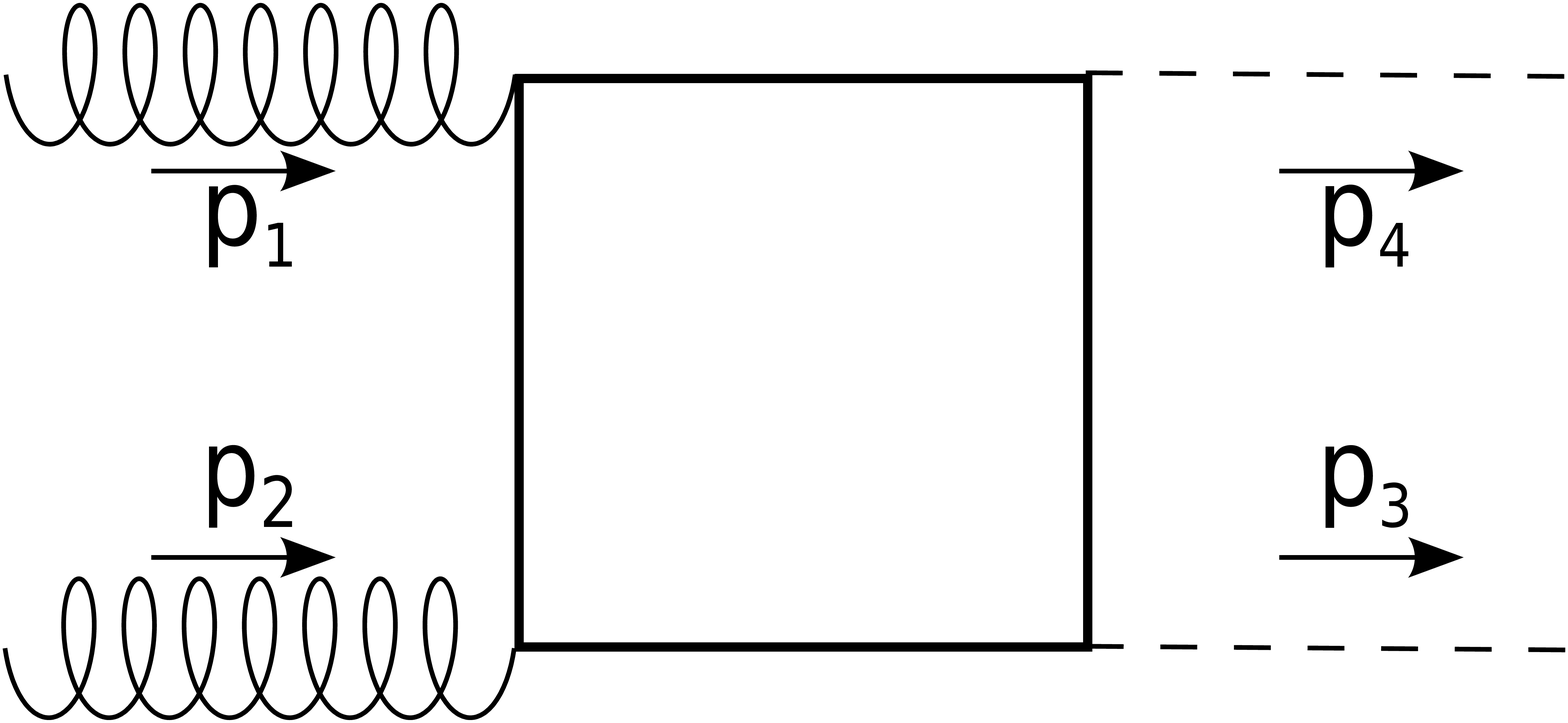}
  \figcaption{\label{boxhh}One-loop box diagram for Higgs pair production via gluon fusion, 
where the initial state momenta are incoming and the final state momenta are outgoing.}
\end{center}

The initial states are on-shell gluons $p_1^2=p_2^2=0$, and the final states are on-shell Higgs bosons $p_3^2=p_4^2=M_H^2$.
The Mandelstam variables are defined as $(p_1+p_2)^2=s$ and $(p_2-p_3)^2=t$.
\footnote{
For simplicity, in the following the dimension of scale is set as GeV by default.
}

\begin{figure}[!htbp]
  \centering
  \subfigure[~${\rm Re}(I_A)$
        \label{boxhhre}]
    {\includegraphics[width=250px]{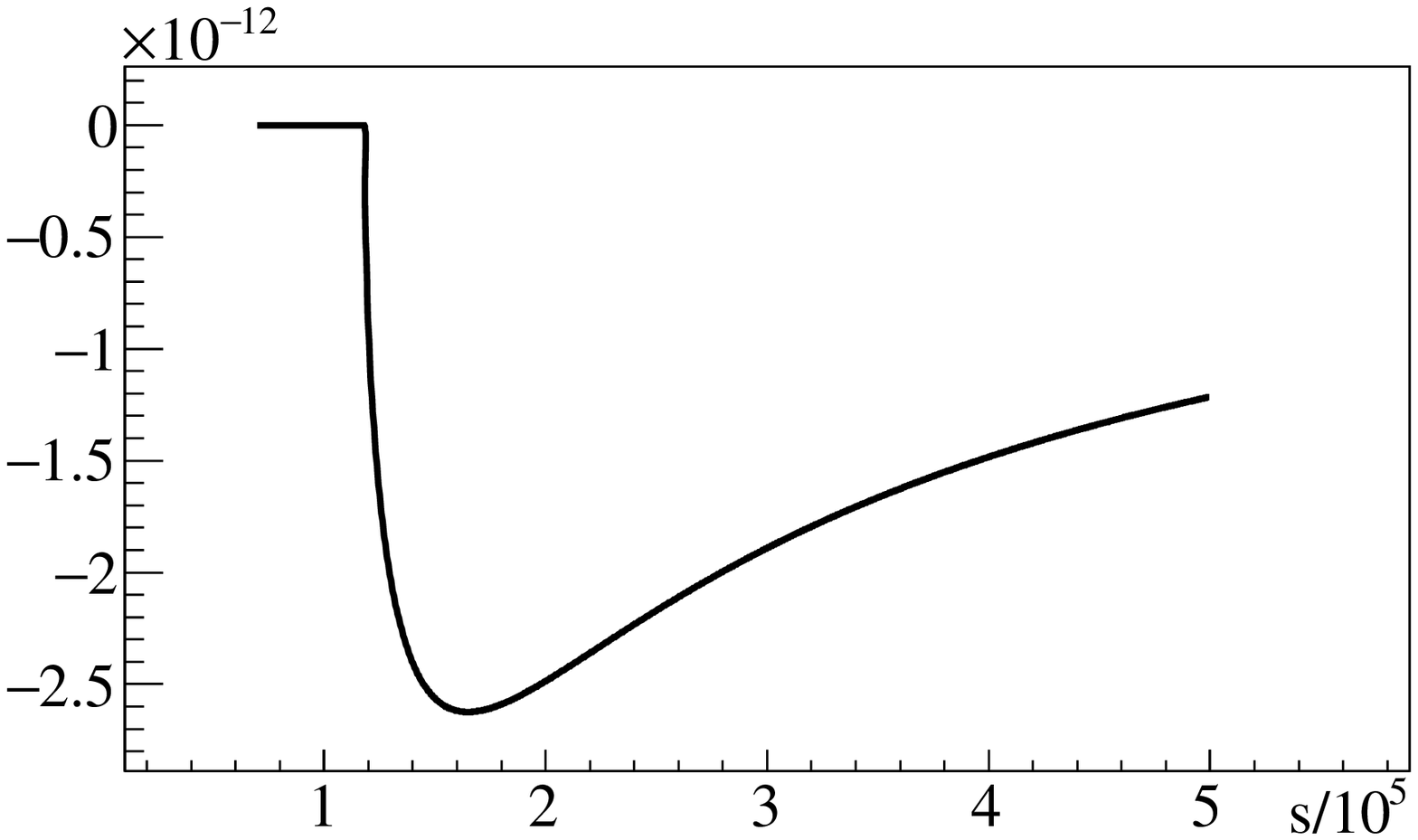}}
  \subfigure[~${\rm Im}(I_A)$
    \label{boxhhim}]
    {\includegraphics[width=250px]{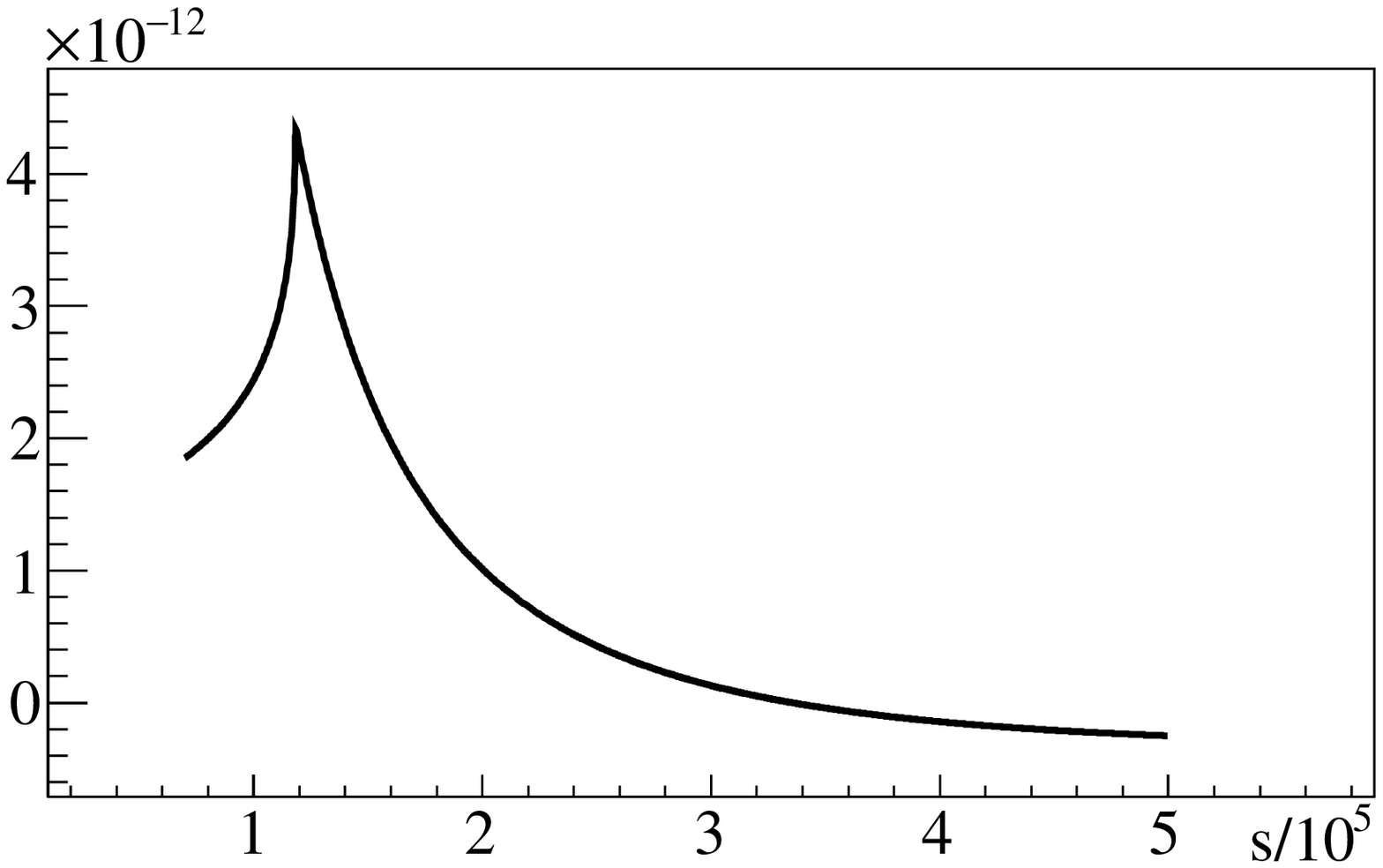}}\\
  \subfigure[~Relative error of ${\rm Re}(I_A)$
        \label{boxhhreer}]
    {\includegraphics[width=250px]{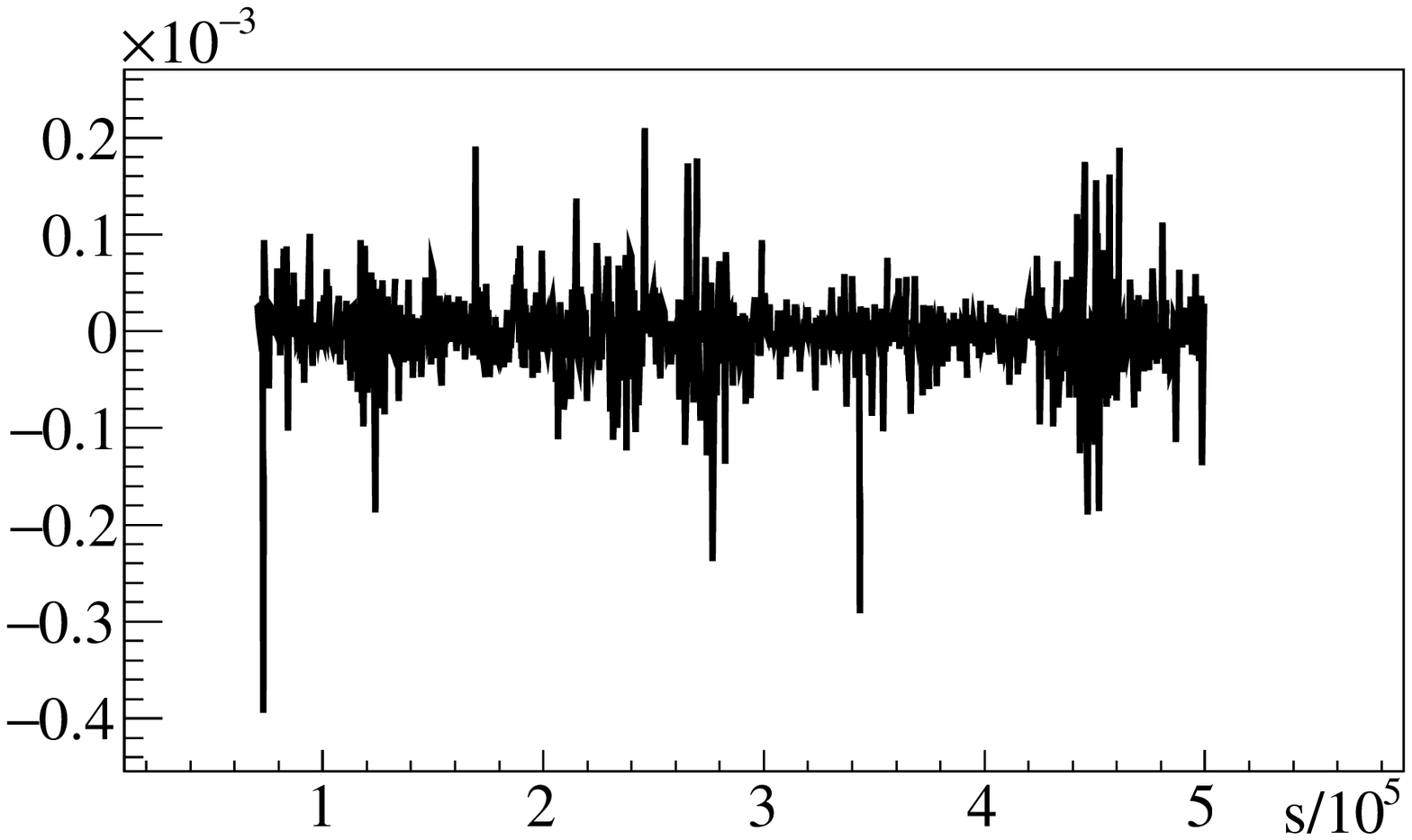}}
  \subfigure[~Relative error of ${\rm Im}(I_A)$
    \label{boxhhimer}]
    {\includegraphics[width=250px]{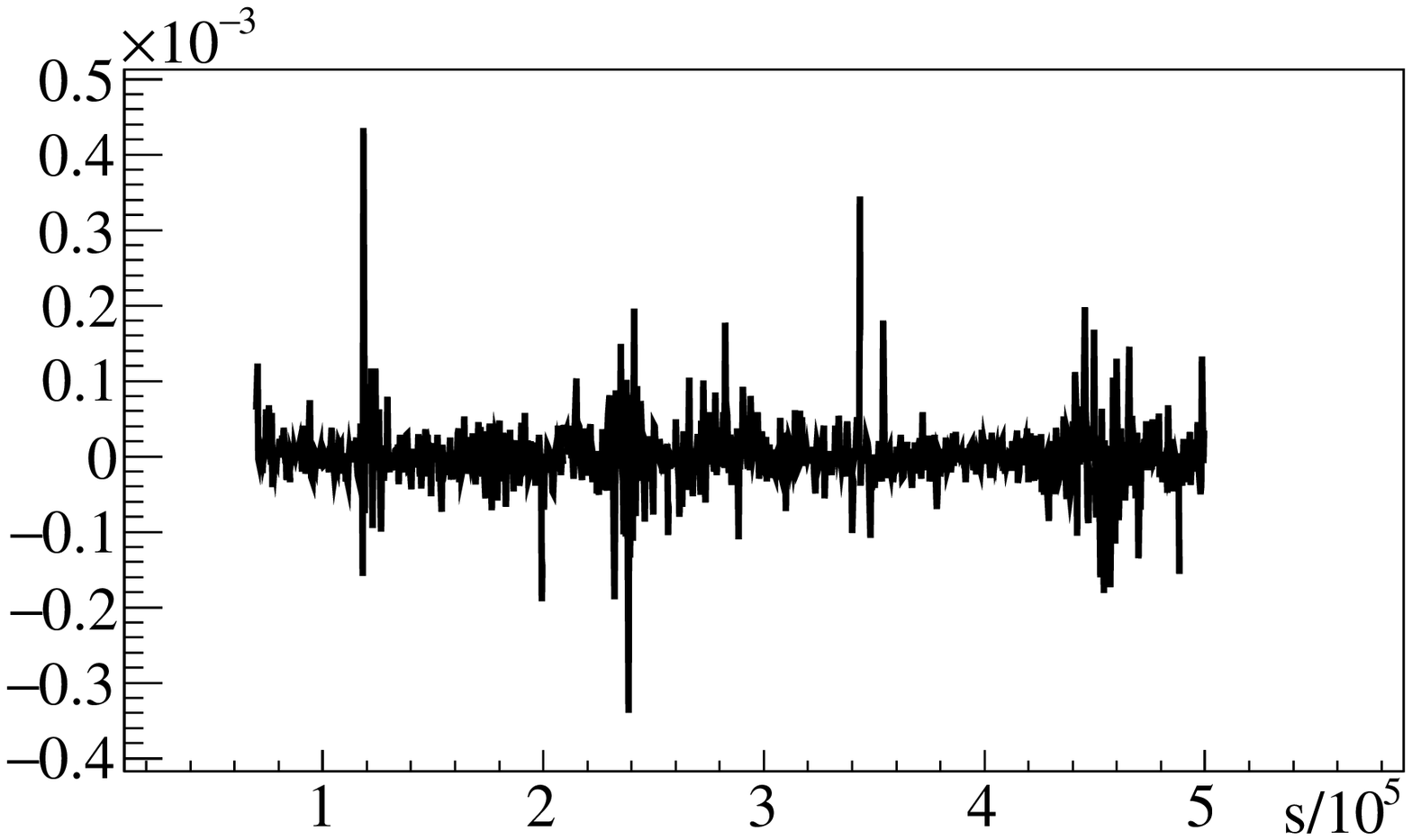}}
\caption{Comparison of single box master integral with LoopTools.}
\label{boxhhfigure}
\end{figure}

As shown in Fig.~\ref{boxhhfigure}, 
we evaluate 1000 points between $s=70000$ and $s=500000$, while $t=-6000$ is fixed.
The average time to evaluate one point is 83 ms, which is acceptable for practical calculation of one-loop Feynman integrals.
We can find that the threshold effect is explicitly shown at $s=4M_t^2=118336$.
Below the threshold ${\rm Re}(I_A)$ vanishes since at the moment the sign of the $F$-term in each decomposed subsector is definite.  
Meanwhile ${\rm Im}(I_A)$ peaks on the threshold, 
where we can see an obvious large relative error due to slow convergence of some integrals.
Fig.~\ref{boxhhfigure} also presents the comparison with results from LoopTools\cite{Hahn:1998yk}.
It can be seen that the relative errors of our results are within $10^{-3}$,
and the relative errors can be smaller than $10^{-4}$ for most of the points.

\end{multicols}
\begin{multicols}{2}
 
\subsection{Two-loop Feynman integral in Euclidean kinematic region\label{loop2Euc}}

\begin{center}
  \includegraphics[width=250px]{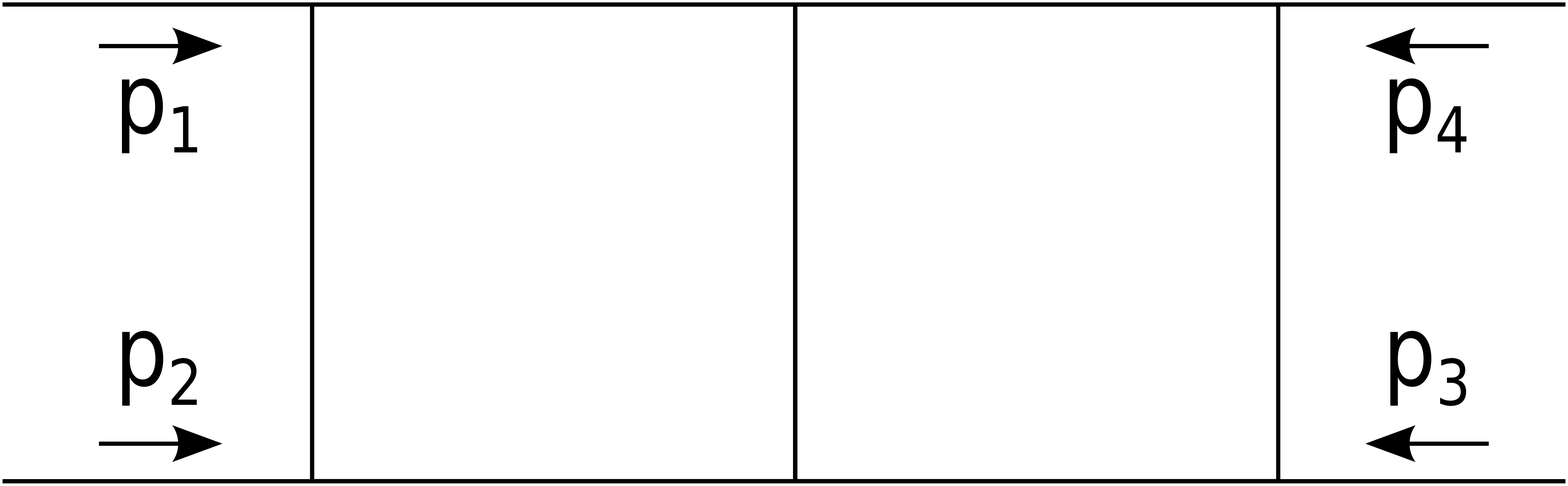}
  \figcaption{Massless double box diagram with two legs off-shell \cite{Binoth:2003ak}, where all the external momenta are incoming.
  \label{bhgp}}
\end{center}

For the demonstration of a two-loop Feynman integral in the Euclidean kinematic region, we choose 
the massless double box diagram with two legs off-shell \cite{Binoth:2003ak}, 
as shown in Fig. \ref{bhgp}. Explicitly this Feynman integral is written as
\begin{align}
\begin{split}
I_B=&\int\dfrac{\mathrm{d}^D k_1}{i\pi^{\dfrac{D}{2}}}\dfrac{\mathrm{d}^D k_2}{i\pi^{\dfrac{D}{2}}}
\dfrac{1}{
(-k_1^2+i\varepsilon)
[-(k_1-p_1)^2+i\varepsilon]
}\\
&\times
\dfrac{1}{
[-(k_1+p_2)^2+i\varepsilon]
[-(k_1-k_2-p_1)^2+i\varepsilon]
}\\
&\times
\dfrac{1}{
(-k_2^2+i\varepsilon)
[-(k_1+p_2-k_2+p_3)^2+i\varepsilon]
}\\
&\times
\dfrac{1}{
[-(k_1-k_2+p_2)^2+i\varepsilon]
}.
\end{split}
\end{align}
Because this Feynman integral contains divergences, the results are expressed in form of a Laurent series in $\epsilon$,
\begin{align}
  I_B=e^{-2\epsilon\gamma_E}s^{-3-2\epsilon}\sum_{i=0}^{i=4}\dfrac{P_i}{\epsilon^i}.
\end{align}
During the numerical evaluation,
the Mandelstam variables are set as $s=(p_1+p_2)^2=-2/3$, $t=(p_2+p_3)^2=-2/3$, and $u=(p_1+p_3)^2=-2/3$.

\end{multicols}
\begin{center}
\tabcaption{Comparison of double box Feynman integral in Euclidean kinematic region.
  \label{gpresult}}
\footnotesize
\begin{tabular}{c*{3}{cc}}
  \toprule
  $(p_1^2,p_2^2,p_3^2,p_4^2)$ & \multicolumn{2}{c}{(-1,0,0,-1)} & \multicolumn{2}{c}{(0,-1,0,-1)} & \multicolumn{2}{c}{(0,0,-1,-1)}\\
  \hline
      & Vegas/CPU & QMC/GPU & Vegas/CPU & QMC/GPU & Vegas/CPU &  QMC/GPU \\
   \hline
$P_4$ & $0.25\pm3\times10^{-6}$ &$0.25\pm1\times10^{-7}$ &  0  & 0 & $0.2501\pm0.0001$ & $0.25\pm2\times10^{-7}$ \\
$P_3$ & $0.40547 \pm0.00004$  &  $0.40546\pm0.00006$ &  0  &  0  & $0.4057 \pm0.005$   & $0.40544\pm0.00003$ \\
$P_2$ & $0.6500\pm0.0003$  & $0.6489\pm0.0005$  & $1.0582 \pm0.0001$ & $1.05799\pm0.00003$ & $3.118  \pm0.002$  & $3.118\pm0.001$ \\
$P_1$ & $-1.183\pm0.001$ & $-1.1823\pm0.0006$ & $1.0938\pm0.0005$ & $1.0947\pm0.0009$ &$12.522 \pm0.007$ & $12.533\pm0.007$ \\
$P_0$ & $-8.798\pm0.004$ & $-8.801\pm0.005$ & $-3.000\pm0.001$  & $-3.003\pm0.002$  &$35.60  \pm0.03$  & $35.60\pm0.03$ \\
\hline
Integration Time & 500s   & 2.2s   &   45s  & 0.84s   &  117s   & 2.2s \\
\bottomrule
\end{tabular}
\end{center}
\begin{multicols}{2}

In Table \ref{gpresult}, we compare our results (marked as QMC) with those from FIESTA3 \cite{Smirnov:2013eza}.
It can be seen that all the results are consistent to $\mathcal{O}(10^{-3})$, 
while our program is much (about 50-200 times) faster than FIESTA3.
This implies that our program can provide the results of Feynman integrals with much higher accuracy in the Euclidean kinematic region.
One of the reasons for the improvement is the QMC algorithm, which has been proved to have
better convergence rate compared to conventional MC algorithm \cite{Kuo2003301,Dick2004493}. Another reason is the implementation of
the GPU-accelerated computing. A GPU has a specified parallel architecture consisting of thousands of smaller,
more efficient cores designed for handling multiple tasks simultaneously. In contrast, a CPU consists of a few cores
optimized for sequential serial processing. The combination of the QMC algorithm and GPU-accelerated computing
provide an efficient way to evaluate the Feynman integrals.

\subsection{Two-loop Feynman integral in physical kinematic region\label{loop2Phy}}

\begin{center}
\includegraphics[width=250px]{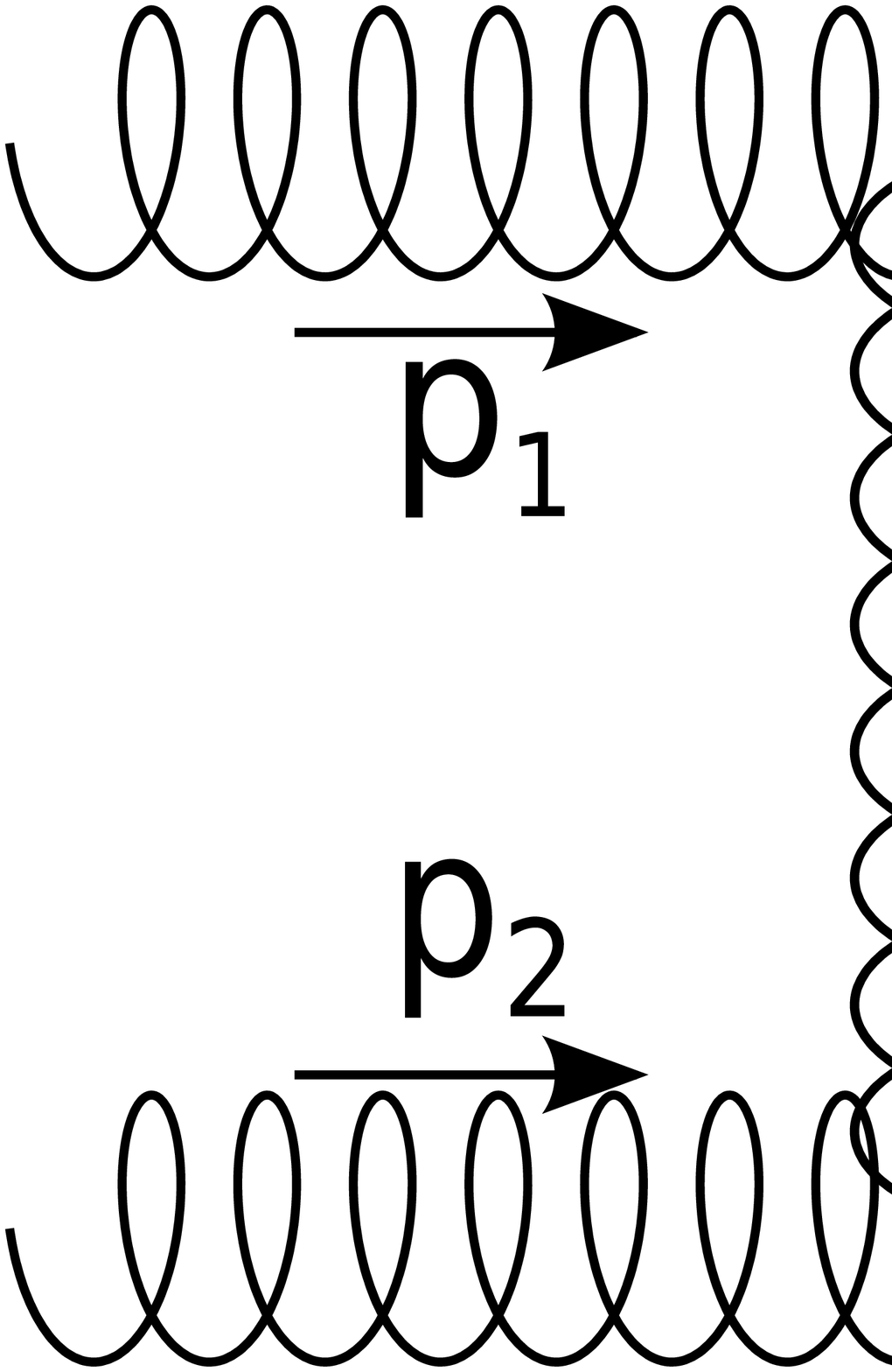}
\figcaption{Two-loop double box Feynman diagram for Higgs pair production via gluon fusion, 
where the initial state momenta are incoming and the final state momenta are outgoing.
\label{gphh}}
\end{center}

The next-to-leading order (NLO) contribution to Higgs pair production via gluon fusion 
contains the two-loop double box Feynman diagram as shown in Fig. \ref{gphh}, 
which is one of the challenges for an analytical approach with a finite top quark mass.
To evaluate this diagram, we will confront the complicated master integral
\begin{align}
\begin{split}
I_C=&\int\dfrac{\mathrm{d}^D k_1}{i\pi^{\dfrac{D}{2}}}\dfrac{\mathrm{d}^D k_2}{i\pi^{\dfrac{D}{2}}}
\dfrac{1}{
(-k_1^2+i\varepsilon)
[-(k_1-p_1)^2+i\varepsilon]
}\\
&\times\dfrac{1}{
[-(k_1+p_2)^2+i\varepsilon]
[-(k_1-k_2-p_1)^2+M_t^2+i\varepsilon]
}\\
&\times\dfrac{1}{
(-k_2^2+M_t^2+i\varepsilon)
[-(k_1+p_2-k_2-p_3)^2+M_t^2+i\varepsilon]
}\\
&\times\dfrac{1}{
[-(k_1-k_2+p_2)^2+M_t^2+i\varepsilon]
}.
\end{split}
\end{align}
This Feynman integral contains IR divergences, so we express the results in form of a Laurent series in $\epsilon$,
\begin{align}
  I_C=e^{-2\epsilon\gamma_E}s^{-3-2\epsilon}\sum_{i=0}^{i=2}\dfrac{P_i}{\epsilon^i}.
\end{align}
Here the initial states are on-shell gluons $p_1^2=p_2^2=0$, and the final states are on-shell Higgs bosons $p_3^2=p_4^2=M_H^2$.
The Mandelstam variables are set as $s=(p_1+p_2)^2=160000$ and $t=(p_2-p_3)^2=-6000$.

\end{multicols}
\begin{center}
\tabcaption{Comparison of two-loop double box Feynman integral in physical kinematic region.
  \label{gphhresult}}
  \footnotesize
\begin{tabular}{ccc}
  \toprule
         &  Vegas/CPU & QMC/GPU \\
                 \hline
  $P_2$  & $-7.959\pm0.009 - 10.586i\pm0.009i $& $-7.949\pm0.003 -10.585i\pm0.005i$ \\
  $P_1$  & $3.9\pm0.1-28.1i\pm0.1i$ & $3.831\pm0.005 -28.022i\pm0.005i$ \\
  $P_0$  & $-3.9\pm0.8+92.3i\pm0.8i$ & $-4.63\pm0.07 + 92.13i\pm0.07i $\\
  \hline
  Integration Time   & 45540s & 19s \\
  \bottomrule
\end{tabular}
\end{center}
\begin{multicols}{2}

As shown in Table \ref{gphhresult}, after more than 12 hours, 
the relative error of the finite term from FIESTA3 can only reach $10^{-2}$. 
By comparison, in 19 seconds our program can obtain an accuracy of about $10^{-3}$,
which may cost FIESTA3 more than a thousand hours.
Moreover, compared with the efficiency improvement obtained in the Euclidean kinematic region in Section \ref{loop2Euc},
we can find a much better improvement in the physical kinematic region  
due to the advantages of the QMC algorithm for complex integrals as explained in Section \ref{QMCalgo}.
Besides, it can be seen that the efficiency of our program for two-loop Feynman integrals
makes the numerical approach viable for the evaluation of the NLO virtual contribution
to Higgs pair production via gluon fusion with a finite top quark mass.
Therefore an investigation of the finite top quark mass effect in Higgs pair production can be accomplished within months.

\subsection{Non-planar Two-loop Feynman integral in physical kinematic region}

\begin{center}
  \includegraphics[width=250px]{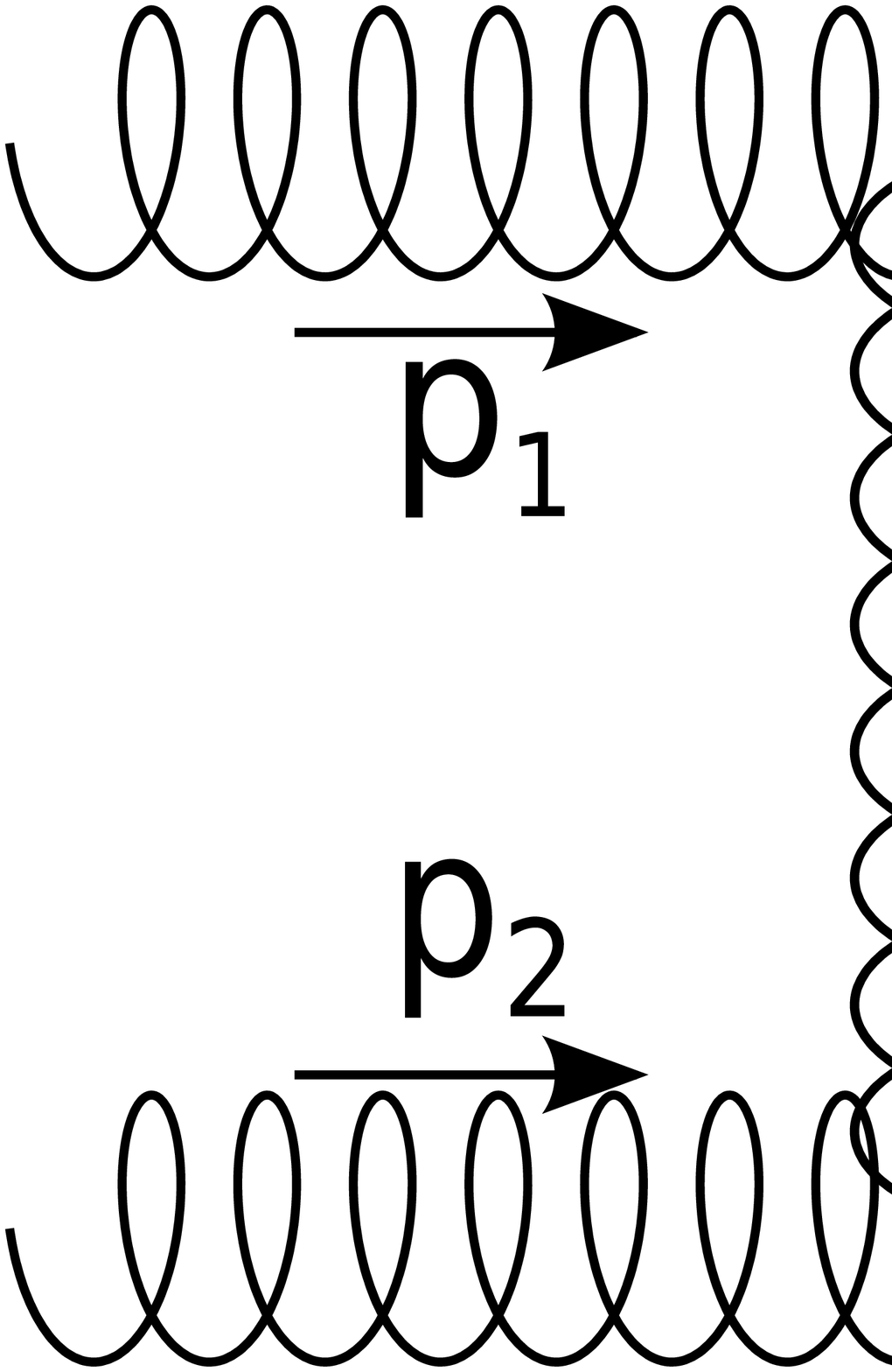}
\figcaption{Non-planar two-loop double box Feynman diagram for the Higgs pair production via gluon fusion, 
where the initial state momenta are incoming and the final state momenta are outgoing.
\label{gnphh}}
\end{center}

The non-planar two-loop diagram shown in Fig. \ref{gnphh} also contributes to Higgs pair production via gluon
fusion at NLO. During the evaluation of this diagram, the most complicated master integral is
\begin{align}
\begin{split}
I_D=&\int\dfrac{\mathrm{d}^D k_1}{i\pi^{\dfrac{D}{2}}}\dfrac{\mathrm{d}^D k_2}{i\pi^{\dfrac{D}{2}}}
\dfrac{1}{
(-k_1^2+i\varepsilon)
[-(k_1+p_1)^2+i\varepsilon]
}\\
&\times\dfrac{1}{
[-(k_1-p_2)^2+i\varepsilon]
[-(k_1+p_1-k_2)^2+M_t^2+i\varepsilon]
}\\
&\times\dfrac{1}{
(-k_2^2+M_t^2+i\varepsilon)
[-(k_2-p_4)^2+M_t^2+i\varepsilon]
}\\
&\times\dfrac{1}{
[-(k_1+p_1-k_2-p_3)^2+M_t^2+i\varepsilon]
},
\end{split}
\end{align}
which contains IR divergences, and can be expressed in form of a Laurent series in $\epsilon$,
\begin{align}
  I_D=e^{-2\epsilon\gamma_E}s^{-3-2\epsilon}\sum_{i=0}^{i=2}\dfrac{P_i}{\epsilon^i}.
\end{align}
The same as the configuration in Section \ref{loop2Phy},
the initial states are on-shell gluons $p_1^2=p_2^2=0$, and the final states are on-shell Higgs bosons $p_3^2=p_4^2=M_H^2$.
The Mandelstam variables are set as $s=(p_1+p_2)^2=160000$ and $t=(p_2-p_3)^2=-6000$.
\end{multicols}
\begin{center}
\tabcaption{Comparison of non-planar two-loop double box Feynman integral in physical kinematic region.
  \label{gnphhresult}}
  \footnotesize
\begin{tabular}{ccc}
  \toprule
         &  Vegas/CPU & QMC/GPU \\
                 \hline
  $P_2$  & $-3.848\pm0.004 + 0.0005i\pm0.003i $ & $-3.8482\pm0.0007 + 0.0004i\pm0.0003i $ \\
  $P_1$  & $3.81\pm0.03-6.41i\pm0.03i $ & $3.83\pm0.02 - 6.40i\pm0.02i $ \\
  $P_0$  & $77.2\pm0.2+20.1i\pm0.2i $ & $77.2\pm0.1 + 19.9i\pm0.1i $\\
  \hline
  Integration Time   & 54290s & 20s \\
  \bottomrule
\end{tabular}
\end{center}
\begin{multicols}{2}

In Table \ref{gnphhresult}, we can see that our program can obtain a result with $\mathcal{O}(10^{-3})$ accuracy in 20 seconds, 
while the relative error of the FIESTA3 result takes over 15 hours to reach near that order of accuracy.
By comparing with the results in previous section, it is obvious that the non-planar double-box master integral
has a slower convergence rate. This is consistent with the conventional conclusion that the evaluation of 
non-planar diagrams is more difficult than planar ones.
Nonetheless, it can be seen that efficiency of our program for the evaluation of non-planar master integrals 
is acceptable for the practical numerical approach, which can provide NLO numerical results for Higgs pair production within months.

\section{Conclusion}
We have implemented the shifted R1LR QMC method associated with CUDA/GPU 
to numerically evaluate Feynman loop integrals by using a sector decomposition algorithm.
Some examples are presented to show the promising efficiency of our program
on the numerical evaluation of Feynman loop integrals. 
For a one-loop box Feynman integral, we could obtain an accuracy of about $10^{-4}$ in tens of milliseconds.
For two-loop double box Feynman integrals, the accuracy can reach about $10^{-3}$ in several seconds 
in the Euclidean kinematic region, while in the physical kinematic region less than half a minute is needed.
The efficiency of our program can make the direct numerical approach viable 
for the precise investigation on some important processes, 
e.g. Higgs pair production via gluon fusion at NLO with the finite top quark mass effect.

\vspace{14mm}

\end{multicols}

\vspace{-1mm}
\centerline{\rule{80mm}{0.1pt}}
\vspace{2mm}

\begin{multicols}{2}
\bibliographystyle{cpc}
\bibliography{qmcmi}

\begin{thebibliography}{10}

\bibitem{Aad:2012tfa}
G.~Aad et~al., Phys. Lett. B, \textbf{716}: 1--29 (2012)

\bibitem{Chatrchyan:2012ufa}
S.~Chatrchyan et~al., Phys. Lett. B, \textbf{716}: 30--61 (2012)

\bibitem{Anzai:2015wma}
C.~Anzai, A.~Hasselhuhn, M.~H{\"o}schele et~al., JHEP, \textbf{07}: 140 (2015)

\bibitem{Anastasiou:2015ema}
C.~Anastasiou, C.~Duhr, F.~Dulat et~al., Phys. Rev. Lett., \textbf{114}(21):
  212001 (2015)

\bibitem{deFlorian:2013jea}
D.~de~Florian and J.~Mazzitelli, Phys. Rev. Lett., \textbf{111}: 201801 (2013)

\bibitem{Chen:2014gva}
X.~Chen, T.~Gehrmann, E.~Glover et~al., Phys. Lett. B, \textbf{740}: 147--150
  (2015)

\bibitem{Boughezal:2015dra}
R.~Boughezal, F.~Caola, K.~Melnikov et~al., Phys. Rev. Lett., \textbf{115}(8):
  082003 (2015)

\bibitem{Boughezal:2015aha}
R.~Boughezal, C.~Focke, W.~Giele et~al., Phys. Lett. B, \textbf{748}: 5--8
  (2015)

\bibitem{Brein:2003wg}
O.~Brein, A.~Djouadi, and R.~Harlander, Phys. Lett. B, \textbf{579}: 149--156
  (2004)

\bibitem{Brein:2011vx}
O.~Brein, R.~Harlander, M.~Wiesemann et~al., Eur. Phys. J. C, \textbf{72}: 1868
  (2012)

\bibitem{Ferrera:2011bk}
G.~Ferrera, M.~Grazzini, and F.~Tramontano, Phys. Rev. Lett., \textbf{107}:
  152003 (2011)

\bibitem{Ferrera:2014lca}
G.~Ferrera, M.~Grazzini, and F.~Tramontano, Phys. Lett. B, \textbf{740}: 51--55
  (2015)

\bibitem{Binoth:2000ps}
T.~Binoth and G.~Heinrich, Nucl. Phys. B, \textbf{585}: 741--759 (2000)

\bibitem{Binoth:2003ak}
T.~Binoth and G.~Heinrich, Nucl. Phys. B, \textbf{680}: 375--388 (2004)

\bibitem{Heinrich:2004iq}
G.~Heinrich and V.~A. Smirnov, Phys. Lett. B, \textbf{598}: 55--66 (2004)

\bibitem{Nagy:2003qn}
Z.~Nagy and D.~E. Soper, JHEP, \textbf{0309}: 055 (2003)

\bibitem{Nagy:2006xy}
Z.~Nagy and D.~E. Soper, Phys. Rev. D, \textbf{74}: 093006 (2006)

\bibitem{Binoth:2005ff}
T.~Binoth, J.~P. Guillet, G.~Heinrich et~al., JHEP, \textbf{10}: 015 (2005)

\bibitem{Bogner:2007cr}
C.~Bogner and S.~Weinzierl, Comput. Phys. Commun., \textbf{178}: 596--610
  (2008)

\bibitem{Smirnov:2013eza}
A.~V. Smirnov, Comput. Phys. Commun., \textbf{185}: 2090--2100 (2014)

\bibitem{Borowka:2015mxa}
S.~Borowka, G.~Heinrich, S.~P. Jones et~al., Comput. Phys. Commun.,
  \textbf{196}: 470--491 (2015)

\bibitem{Lepage:1977sw}
G.~P. Lepage, J. Comput. Phys., \textbf{27}: 192 (1978)

\bibitem{Bogner:2010kv}
C.~Bogner and S.~Weinzierl, Int. J. Mod. Phys. A, \textbf{25}: 2585--2618
  (2010)

\bibitem{Heinrich:2008si}
G.~Heinrich, Int. J. Mod. Phys. A, \textbf{23}: 1457--1486 (2008)

\bibitem{Kaneko:2009qx}
T.~Kaneko and T.~Ueda, Comput. Phys. Commun., \textbf{181}: 1352--1361 (2010)

\bibitem{Kaneko:2010kj}
T.~Kaneko and T.~Ueda, {Sector Decomposition Via Computational Geometry}, in
  \emph{{Proceedings of 13th International Workshop on Advanced computing and
  analysis techniques in physics research (ACAT2010)}}, p. 082

\bibitem{Anastasiou:2007qb}
C.~Anastasiou, S.~Beerli, and A.~Daleo, JHEP, \textbf{0705}: 071 (2007)

\bibitem{ANU:8877392}
J.~Dick, F.~Y. Kuo, and I.~H. Sloan, Acta Numerica, \textbf{22}: 133--288
  (2013)

\bibitem{Kuo2003301}
F.~Kuo, Journal of Complexity, \textbf{19}(3): 301 -- 320 (2003), oberwolfach
  Special Issue

\bibitem{Dick2004493}
J.~Dick, Journal of Complexity, \textbf{20}(4): 493 -- 522 (2004)

\bibitem{Agashe:2014kda}
K.~A. Olive et~al., Chin. Phys. C, \textbf{38}: 090001 (2014)

\bibitem{Hahn:2004fe}
T.~Hahn, Comput. Phys. Commun., \textbf{168}: 78--95 (2005)

\bibitem{Hahn:1998yk}
T.~Hahn and M.~Perez-Victoria, Comput. Phys. Commun., \textbf{118}: 153--165
  (1999)

\end{thebibliography}

\end{multicols}


\end{document}